\documentclass[sigconf]{acmart}

\AtBeginDocument{%
  }

\copyrightyear{2024}
\acmYear{2024}
\setcopyright{rightsretained}
\acmConference[RecSys '24]{18th ACM Conference on Recommender Systems}{October 14--18, 2024}{Bari, Italy}
\acmBooktitle{18th ACM Conference on Recommender Systems (RecSys '24), October 14--18, 2024, Bari, Italy}\acmDOI{10.1145/3640457.3688130}
\acmISBN{979-8-4007-0505-2/24/10}

\usepackage{bbm}
\usepackage{color}
\usepackage{algorithm}
\usepackage{algorithmic}
\usepackage{amsmath}
\renewcommand{\algorithmicrequire}{\textbf{Input:}}
\renewcommand{\algorithmicensure}{\textbf{Output:}}
\usepackage{bm} 
\usepackage{graphicx}

\begin{document}

\title{Fair Reciprocal Recommendation in Matching Markets}

\author{Yoji Tomita}
\affiliation{%
  \institution{CyberAgent, Inc.}
  \city{Tokyo}
  \country{Japan}
}
\email{tomita_yoji@cyberagent.co.jp}

\author{Tomohiko Yokoyama}
\affiliation{%
  \institution{The University of Tokyo}
  \city{Tokyo}
  \country{Japan}
}
\email{tomohiko_yokoyama@mist.i.u-tokyo.ac.jp}

\renewcommand{\shortauthors}{Tomita and Yokoyama}

\begin{abstract}
    Recommender systems play an increasingly crucial role in shaping people's opportunities, particularly in online dating platforms.
    It is essential from the user's perspective to increase the probability of matching with a suitable partner while ensuring an appropriate level of fairness in the matching opportunities.
    
    We investigate reciprocal recommendation in two-sided matching markets between agents divided into two sides. In our model, a match is considered successful only when both individuals express interest in each other. Additionally, we assume that agents prefer to appear prominently in the recommendation lists presented to those on the other side.
    We define each agent's \emph{opportunity} to be recommended and introduce its fairness criterion, \emph{envy-freeness}, from the perspective of fair division theory. 
    The recommendations that approximately maximize the expected number of matches, empirically obtained by heuristic algorithms, are likely to result in significant unfairness of opportunity. Therefore, there can be a trade-off between maximizing the expected matches and ensuring fairness of opportunity.
    To address this challenge, we propose a method to find a policy that is close to being envy-free by leveraging the \emph{Nash social welfare} function. Experiments on synthetic and real-world datasets demonstrate the effectiveness of our approach in achieving both relatively high expected matches and fairness for opportunities of both sides in reciprocal recommender systems.
\end{abstract}






\maketitle

\section{Introduction}

In recent decades, online matching platforms such as those for dating and job recommendations have proliferated rapidly across many countries.
In these platforms, recommender systems play a crucial role in the success of services. 
Unlike conventional recommender systems (RSs) that suggest items (such as movies or music) to users, recommender systems in online matching platforms recommend users from one side of the platform to users on the other side, and vice versa. 
These systems have been developed recently as ``reciprocal recommender systems (RRSs)''~\cite{pizzato2010recon,mine2013reciprocal,palomares2021reciprocal}.

In many online matching platforms, one of the most important objectives is to maximize the number of matches between users.
To increase matches in a service, it is essential to avoid the concentration of recommendation opportunities on popular users.
This is because even if a highly popular user is recommended to many users on the other side and receives numerous expressions of interest, the number of connections they can meaningfully engage with is physically limited.
Several recent works show that practical online dating recommendation algorithms suffer from this popularity bias~\cite{chen2023reducing,celdir2024popularity}, therefore some reciprocal recommendation algorithms have been developed to avoid popularity bias and increase the total number of matches~\cite{chen2023reducing,tomita2022matching,Tomita2023,su2022optimizing}.
However, avoiding the concentration of recommendation opportunities on popular users and distributing them more evenly is also important in terms of fairness among users. This fairness consideration adds another layer of complexity to the design of effective RRSs.

In the literature on conventional recommendation systems, fairness issues have recently attracted increased attention~\cite{Li2022Fairness,Wang2023FairnessRecommender,DeldjooJannach2024}, and studied from various aspects such as user-side~\cite{ekstrand2018all,wang2021user}, item-side~\cite{beutel2019fairness,zhu2020measuring}, both-side~\cite{WuCaoXuTan2021},
group-fairness~\cite{kearns2018preventing,kearns2019empirical}, and individual-fairness~\cite{biega2018equity,singh2018fairness}.
On the other hand, studies on fairness issues of RRSs in online matching platforms have been limited~\cite{xia2019we,Virginie2021}.
However, we argue that fairness issues are crucial especially for RRSs, and more research is needed from real-world platforms.
Unless the conventional RSs in which providers of recommended items (movies, music, consumer goods) are firms, both users who receive recommendation lists and those who are recommended to other users are individual people in online matching platforms such as online dating, and thus unfairness among users directly affects reputations of the service.
Moreover, since each user acts both as an active user who receives recommendation lists and clicks (to send ``likes'' or apply) and as a passive user who is recommended to other side of users in online matching platforms, the structures of the platforms are entirely different from the conventional RSs from perspective of users, and thus the fairness in RRSs needs different definition and analysis.

In this paper, we consider a new notion of fairness in RRSs in online matching platforms by ``envy-freeness'', which has been studied in the fair division in economics and algorithmic game theory~\cite{Foley1967,Varian1974,Budish2011}.
In the fair division, which studies how to allocate divisible or indivisible goods fairly among people, an agent $i$ is said to have \emph{envy} toward another agent $i'$ if the agent $i$ prefers the allocation for $i'$ to his own allocation, and it is said that an allocation is \emph{envy-free} if no agent has envy in a given allocation.
\citet{SaitoJoachimsKDD2022} apply the concept of fairness based on envy-freeness to item-side fairness in conventional RSs, and they propose recommendation policies maximizing \emph{Nash social welfare (NSW)} that achieve approximately envy-free recommendation among recommended items.
Since users' recommendation opportunities, which is defined as how frequently one user is recommended to the other side affects directly the users' satisfaction from the service, fairness of recommendation opportunities is essential in the online matching platforms.
Moreover, as shown in the latter sections, we can interpret the design of recommendation in RRSs as the allocation of the recommendation opportunities among users, and thus it is natural to consider fairness based on envy-freeness from the fair division theory.
However, the problem of RRSs on online matching platforms is different from conventional RSs, and the extension of the definition of envy-freeness and methods of \citet{SaitoJoachimsKDD2022} to RRSs raises a problem.

The contribution of this paper is as follows.
We first formalize the problem of RRSs in the online matching platforms and define fairness concepts of recommendation opportunities among users based on envy-freeness, which is an application of the fair division theory and \citet{SaitoJoachimsKDD2022} to the problem of RRSs.
We next describe the social welfare (SW) maximization method that maximizes the expected number of matches and show that it causes a non-negligible unfairness among users and thus there is a trade-off between the number of matches and fairness by simplified example and numerical experiments.
Finally, we propose the alternate NSW maximization method via Frank-Wolfe which achieves approximately fair recommendation among users in RRSs, and show that it gives a competitive level of total matches rather than existing methods and almost zero envies of recommendation opportunities in experiments with synthetic data and real-world data.

\subsection{Related Work}

\paragraph{Reciprocal Recommender Systems (RRSs)}

Recently, RRSs have attracted increased attention and are surveyed in a comprehensive way by \citet{palomares2021reciprocal}.
However, most of the previous studies in this field focused on estimation of preference scores of users towards other side of users~\cite{pizzato2010recon,xia2015reciprocal,neve2019aggregation,neve2019latent}.
Some recent works have studied the post-processing phase to compute reciprocal scores of user pairs.
\citet{su2022optimizing} formalize the problem of the post-processing phase in reciprocal recommender systems and propose a new algorithm based on convex programming.
\citet{Tomita2023} study the similar model of \citet{su2022optimizing} and consider the TU matching algorithm from the matching theory in economics, whose application in online dating is also studied by \citet{tomita2022matching} and \citet{chen2023reducing}.
They also studied methods to avoid the concentration of recommendation on popular users.
%to maximize the overall matches. 
However, they did not consider directly fairness issues among users in RRSs.

\paragraph{Fair Reciprocal Recommendation}
Although not numerous, there are a few notions of fairness that are relevant to RSSs. \citet{xia2019we} discuss fairness among groups and similarity of mutual preference in RSSs based on Walrasian equilibrium, which is an economic concept considering a balance between supply and demand while optimally satisfying the preferences of both parties involved. 
\citet{Virginie2021} consider fairness in RRSs based on Lorenz dominance. They define fair rankings as those with non-dominated Lorenz curves for both users and items.

These model differ significantly from ours, as it does not explicitly require mutual interest between users for a successful match, which is crucial in our RRS.

\paragraph{Fair Division in Machine Learning}
The fair division theory and envy-freeness have gained attentions in the machine learning, especially, classifications and recommendations~\cite{balcan2019envyfree,golz2019paradoxes}.
In the context of binary classification, \citet{balcan2019envyfree} propose envy-freeness as a notion of fairness for classification tasks where individuals have diverse preferences over the possible outcomes.
We mentioned several studies on a fair recommendation systems. Here, we highlight a few notable works that have adopted envy-freeness to ensure fairness in recommendation settings.
\citet{patro2020fairrec} explore fairness in product recommendations to customers, considering both customer utility and producer exposure.
They based their fairness metrics on envy-freeness and maximin share (MMS), which is another fairness metric, and proposed fair recommendation algorithms utilizing fair division techniques.
\citet{do2022online} study fairness in user-item recommendations, focusing on envy-freeness among users. They proposed a method for certifying the fairness of the recommendation system based on envy-freeness criteria.

\paragraph{Fair Division for Indivisible Goods}
The problem of fair division has been extensively studied in various fields~\cite{steinhaus1948problem,Foley1967,Varian1974,Budish2011}.
It concerns allocating resources to agents in a fair and efficient manner.
In recent years, there has been active research on fair division for indivisible items, i.e., goods can not be shared between agents. In such cases, envy-free allocations may not exist, and a relaxed notion of envy-freeness, \emph{envy-freeness up to one good (EF1)}, have been studied~\cite{Budish2011,Lipton2004}.
An allocation is EF1 if envies can be eliminated by removing at most one good from the subset of items which an envied agent receives.
Another key concept in fair division is Nash social welfare (NSW), which aims to maximize the product of agents' utilities, striking a balance between efficiency and fairness. 
When goods are divisible and agents have additive valuations, any maximum Nash social welfare (MNW) allocation is both envy-free and \emph{Pareto optimal (PO)}~\cite{Eisenberg1959,Varian1974}, where PO ensures that no other allocation makes an agent better off without making someone else worse off. 
When goods are indivisible and agents' valuations are additive, \citet{Caragiannis2019} showed that any MNW allocation is both EF1 and PO. However, computing MNW is NP-hard~\cite{Nguyen2014,Lee2017}. A substantial body of research has been devoted to developing approximation algorithms aimed at maximizing the NSW~\cite{ColeGkatzelis2018,Garg2023,JainVaish2024}.
We employ these concepts of fairness and efficiency in the context of RRSs.

Recently, some papers have combined fairness concepts in fair division with two-sided matching markets~\cite{freeman2021two,gollapudi2020almost,igarashi2023fair}.
Our reciprocal recommendation problem can be considered to include the many-to-many matching problem when the examination function is limited to uniform and only deterministic recommendation list is considered. 
\citet{freeman2021two} proposed the notion of \emph{double envy-freeness up to one match (DEF1)} in two-sided matching markets, which simultaneously satisfies fairness on both sides.
They show that, when both sides have identical ordinal preferences, a complete matching satisfying DEF1 always exists and can be computed in polynomial time.
%They focus on the matching problem, not in the context of recommendations. 
Our model assumes that agents apply to other agents with probabilities dependent on the ranking position based on the given recommendations, which makes our setting significantly different from theirs.

\section{Preliminaries}

We consider the recommendation process of two-sided matching platforms, such as those used in online dating services.
Mirroring the approach of major online dating platforms in the real world, users on one side of the market receive recommendation lists of users from the other side. 
Users then choose which users from the opposite side to apply to (or send ``likes''). This process operates reciprocally, with users on both sides receiving recommendations and making applications.
Then,
a match between two users from opposite sides is established through a process of mutual interest. 

Although the primary objective of the platform is to maximize the number of matches overall, we argue that fairness among users raises a problem.
In this section, we first explain our model of this platform and define a notion of fairness inspired by the fair division theory.

\subsection{Model}
In our model, we consider matching markets comprised of two distinct sets of \emph{agents}. 
Let $[k] = \{1,2,\ldots,k\}$ for a positive integer $k$.
We define $N=\{a_1,a_2,\ldots,a_n\}$ as the set of $n$ agents on the left side and $M=\{b_1,b_2,\ldots,b_m\}$ as the set of $m$ agents on the right side of the market.
For each pair of agents $(a_i, b_j)$ where $i\in [n]$ and $j\in [m]$, we introduce estimated \emph{preference probabilities}.
Let $\hat{p}_1(i,j)\in [0,1]$ represent the estimated preference probability of agent $a_i\in N$ for agent $b_j\in M$.
This probability indicates the likelihood that agent $a_i \in N$ prefers or would choose agent $b_j \in M$. 
Similarly, let $\hat{p}_2(j,i)\in [0,1]$ denote the estimated preference probability of agent $b_j\in M$ for agent $a_i\in N$.
It is important to note that these preference probabilities are typically derived from data-driven methods and are assumed to be given in our framework.
Furthermore, we allow for asymmetric preferences, meaning that $\hat{p}_1(i,j)$ may not necessarily equal $\hat{p}_2(j,i)$ for any given pair of agents.

In the context of RRSs, each agent receives a \emph{recommendation list} of the agents on the opposite side.
We employ probabilistic recommendation lists represented as doubly stochastic matrices.

For a positive integer $d$, a $d \times d$ matrix $S \in \mathbb{R}^{d \times d}_{\geq 0}$ is called a doubly stochastic matrix if the sum of each row and each column is equal to $1$. 

In our model, each agent $a_i\in N$ receives a doubly stochastic matrix $A_i \in \mathbb{R}^{m\times m}_{\geq 0}$ representing a probabilistic ranking. We refer to $A_i$ as the \emph{recommendation} for agent $a_i\in N$. The $(j,k)$ entry of $A_i$, denoted as $A_i(j,k)$, represents the probability that agent $b_j\in M$ is placed in the $k$-th position in the ranking of agent $a_i$.
Similarly, agent $b_j\in M$ receives a doubly stochastic matrix $B_j \in \mathbb{R}^{n\times n}_{\geq 0}$ and we call $B_j$ the recommendation for agent $b_j\in M$. 
We denote $\bm{A}=(A_1,A_2,\ldots,A_n)$ and $\bm{B}=(B_1,B_2,\ldots,B_m)$.
We define the \emph{recommendation policy} $\pi$ as the pair $(\bm{A},\bm{B})$, which encompasses the recommendations for all agents.

In this paper, we focus on recommendation policies that aim to maximize the overall utility for all agents on both sides of the market, where utility is derived from matches resulting from the recommendations.
Our primary objective in this paper is to develop a framework for identifying ``optimal'' recommendation policies $\pi$, given the estimated preference probabilities.

Next, we explain how agents apply to agents on the opposite side based on the recommendations they receive.
We adopt the position-based model (PBM), a commonly assumed model in recommendation contexts~\cite{craswell2008experimental,Joachims2017-ix}, to determine the probability of an agent applying to another agent.
Let $v(k)$ denote the examination probability, representing the likelihood that a user examines the $k$-th candidate in their recommendation list.
The function $v$ is non-increasing with respect to $k$, reflecting the observation that users are less likely to consider candidates lower in the list. The specific definition of $v$ depends on the real-world application context.
Common examples of examination probability functions in the literature include $v(k) = \frac{\mathbbm{I}(k\leq K)}{k} $ 
\footnote{Agents will not apply to agents who are beyond the $K$-th position in the ranking, or they may not receive recommendations for such agents. For an event $\mathcal{E}$, $\mathbbm{I}(\mathcal{E})$ is an indicator function such that $\mathbbm{I}(\mathcal{E})=1$ if $\mathcal{E}$ holds and $0$ otherwise.} or $\frac{\mathbbm{I}(k\leq K)}{\log_2 (k+1)}$, where $K \geq 0$ represents a threshold number. These functions are widely used~\cite{jarvelin2002cumulated,SaitoJoachimsKDD2022}.
%{\color{orange}[TODO]cite}.
%
Based on PBM, we suppose that
the probability that an agent $a_i \in N$ applies to an agent $b_j\in M$ of the opposite side is given by the product of $a_i$'s preference for $b_j$ and the expected value of $v(k)$ where $k$ is the position of $b_j$ in the recommendation list for $a_i$.
Thus, given recommendation $A_i$, the probability that agent $a_i\in N$ applies to agent $b_j\in M$ is given by 
$$
    \mathrm{Pr}[\text{$a_i$ applies to $b_j$}] 
    = \hat{p}_1(i,j)\sum_{k=1}^{m} v(k)  \cdot A_i(j,k),
$$  
and similarly the probability that agent $b_j\in M$ applies to agent $a_i\in N$ is given by 
$$
    \mathrm{Pr}[\text{$b_j$ applies to $a_i$} ] 
    = \hat{p}_2(j,i)\sum_{\ell=1}^{n} v (\ell)  \cdot B_j(i,\ell).
$$

In our model, a match between two agents $a_i\in N$ and $b_j \in M$ is established only when $a_i$ applies to $b_j$ and also $b_j$ applies to $a_i$ mutually.
Thus, the probability that the agent $a_i\in N$ matches with the agent $b_j \in M$ is given by
\begin{align*}
    \mathrm{Pr}[\text{$a_i$ matches with $b_j$} ] 
    &=
    \mathrm{Pr}[\text{$b_j$ matches with $a_i$} ]  \\
    &= 
    \mathrm{Pr}[\text{$a_i$ applies to $b_j$} ~\land~ \text{$b_j$ applies to $a_i$}] \\
    &= \mathrm{Pr}[\text{$a_i$ applies to $b_j$} ] \cdot \mathrm{Pr}[\text{$b_j$ applies to $a_i$} ] \\
    &= p_{i,j} \sum_{k=1}^m\sum_{\ell=1}^{n} v_{k\ell}  A_{i}(j,k) \cdot B_{j}(i,\ell),
\end{align*}
where $p_{i,j} = \hat{p}_1(i,j)\cdot \hat{p}_2(j,i)$ and $v_{k\ell} = v(k)\cdot v(\ell)$ for each $k \in [m]$ and $\ell \in [n]$.

In our two-sided recommendation setting, we assume that each agent obtains \emph{utility}, defined as the expected number of successful matches with other agents from the opposite side, based on the recommendation policy.
Note that an individual may match with more than one partner.
Specifically, under policy $\pi$, the utility of agent $a_i\in N$ is given by 
$$
U_{i}(\pi) = \sum_{b\in M} \mathrm{Pr}[\text{$a_i$ matches with $b$}].
$$ Here, the utility $U_{i}(\pi)$ only depends on $A_i$ and $\bm{B}$.
Similarly, the utility of agent $b_j\in M$ under policy $\pi$ is given by 
$$V_{j}(\pi) = \sum_{a\in N} \mathrm{Pr}[\text{$b_j$ matches with $a$}].
$$

We now define the \emph{social welfare} $\mathrm{SW}(\pi)$ for policy $\pi$ as the expected number of matches between all men and all women under $\pi$, that is,
$$
\mathrm{SW}(\pi) 
    =  \sum_{i\in [n]} U_i(\pi) 
    = \sum_{j\in [m]} V_j(\pi).
$$
A policy $\pi$ is said to be \emph{socially optimal} if it maximizes the social welfare.

If an agent is aware of the recommendations of all agents  on the opposite side, the agent would desire the most favorable recommendation to maximize its own utility.
For each agent $a_i \in N$, we say that the recommendation $A_i^*$ is the \emph{best} for given $\bm{B}$ if $A_i^* \in \mathrm{argmax}_{A_i}U_i (A_{i},\bm{B})$, under the constraint that $A_i$ is a doubly stochastic matrix.
Similarly, for agent $b_j\in M$, recommendation $B_j^*$ is best for given $\bm{A}$ if $B_j^* = \mathrm{argmax}_{B_j}U_j (\bm{A},B_{j})$. 
It is straightforward to verify that if the recommendation policy is socially optimal, then all agents receive their best recommendations.

\subsection{Fairness Axioms}
We introduce fairness notions for recommendation policies in our reciprocal recommendation model.

We first define the concept of the \emph{opportunity} to be recommended, inspired by the concept of the \emph{impact} defined by \citet{SaitoJoachimsKDD2022}\footnote{Saito and Joachims~\cite{SaitoJoachimsKDD2022} defined the concept of the impacts for items that are recommended to users, and introduce envy-freeness among items with respect to impacts. 
If we map agents in $N$ to users and agents in $M$ to items and fix $\bm{A}$, then the opportunity of agent $j$ in our model precisely coincides with the impact of item $j$.}.
The opportunity for an agent represents all the rows relevant to that agent from every recommendation list provided to the opposite side.
Specifically, the opportunity of $a_i\in N$ is represented by an opportunity matrix $C_i \in \mathbb{R}^{n \times m}$, defined as
$$ 
C_i = (B_1(i,\cdot)^{\top}, B_2(i,\cdot)^{\top}, \ldots, B_m(i,\cdot)^{\top}), 
$$
where $B_j(i,\cdot)\in \mathbb{R}^{n}$ is the $i$-th row of the recommendation matrix $B_j$ for each $j \in [m]$, and $B_j(i,\cdot)^{\top}$ denotes its transpose.

Similarly, for an agent $b_j \in M$, we define the opportunity matrix $D_j \in \mathbb{R}^{m \times n}$ as
$$
D_j = (A_1(j,\cdot)^{\top}, A_2(j,\cdot)^{\top}, \ldots, A_n(j,\cdot)^{\top}),
$$ 
where $A_i(j,\cdot)\in \mathbb{R}^{m}$ is the $j$-th row of the recommendation matrix $A_i$ for each $i \in [n]$.

Next, we define a concept of fairness for recommendation policy that ensures the opportunities to be recommended are "fair" for all agents from the perspective of fair division theory \cite{Foley1967,Varian1974}.
For instance, consider a scenario where users can assess the probability of being recommended to the other side. 
If two users from the same side have identical situations (e.g., similar features or preference levels from the opposite side), however one user is clearly being recommended more favorably to the other side than the other user, the disadvantaged user may feel envious and dissatisfied with the platform.

To capture such states of unfairness, we introduce the concept of \emph{envy-freeness} with respect to opportunity.
An agent $a_i\in N$ is said to \emph{envy} another agent $a_{i'}\in N$ in terms of opportunity if the utility that agent $a_i$ would receive from $a_{i'}$'s opportunity exceeds the utility that $a_i$ receives from their own opportunity.
Let $U_i\bigl(C_{i'}\bigr)$ denote the utility when agent $a_i$ receives from $a_{i'}$'s opportunity $C_{i'}$. That is,
\[
    U_i\bigl(C_{i'}\bigr) = \sum_{j\in [m]} \sum_{k=1}^{m} \sum_{\ell=1}^{n} p_{i,j} v_{k\ell} A_{i}(j,k)\cdot B_{j}(i',\ell).
\]
Agent $a_i$ envies agent $a_{i'}$ if and only if $U_i\bigl(C_{i}\bigr) < U_i\bigl(C_{i'}\bigr)$.

The situation is analogous for agents in $M$. 
For an opportunity $D_{j'} \in \mathbb{R}^{m\times n}$ of agent $j'\in M$, we define the utility which agent $b_j \in M$ receives from the opportunity $D_{j'}$ as
\[
    U_j\bigl(D_{j'}\bigr) = \sum_{j\in [m]} \sum_{k=1}^{m} \sum_{\ell=1}^{n} p_{i,j} v_{k\ell} A_{i}(j',k)\cdot B_{j}(i,\ell).
\]
Note that we have $U_i (\pi) = U_i\bigl(C_i\bigr)$ for every $i\in [n]$, and $U_j (\pi) = U_j\bigl(D_j\bigr)$ for every $j\in [m]$.

A policy is \emph{double envy-free} if no agent envies any other agent's opportunity on the same side of the market. Formally,
\begin{definition}[Double envy-free policy]
    The opportunity matrices $C_1,C_2,\ldots,C_n$ satisfy
    \emph{left-side envy-freeness} if for every pair of agents $a_{i},a_{i'} \in N$,
        $U_i\bigl(C_i\bigr) \geq U_i\bigl(C_{i'}\bigr)$.
    Similarly, the opportunity matrices $D_1,D_2,\ldots,D_m$ satisfy \emph{right-side envy-freeness} if for every pair of agents $b_{j},b_{j'}\in M$, $U_j\bigl(D_j\bigr) \geq U_j\bigl(D_{j'}\bigr)$.
    A recommendation policy is \emph{double envy-free} if its opportunity matrices of both sides simultaneously satisfy left-side and right-side envy-freeness.
\end{definition}
It is worth noting that a double envy-free policy always exists in RRSs. 
To illustrate this, consider a scenario where we assign probabilistic recommendation lists such that all agents have an equal probability of appearing at any position in the lists of agents on the opposite side.  Formally, we can express this as a recommendation policy $\pi=(\bm{A},\bm{B})$ where
$A_i(j,k)=\frac{1}{m}$ for all $i \in [n], j \in [m], k \in [m]$ and 
$B_j(i,\ell)=\frac{1}{n}$ for all $j \in [m], i \in [n], \ell \in [n]$.
However, it may not necessarily result in a high number of successful matches.

%
%Furthermore, 
Conversely, maximizing the social welfare does not guarantee that the resulting policy will be double envy-free. 
The presence of conflicting preferences among agents can lead to a trade-off between optimizing the overall utility and ensuring fairness for all participants.
Consider the following example.
\begin{example}\label{example:non-envy-free}
    Consider a simple scenario with two agents on $N$ and one agent on $M$. We will use the inverse examination function $v(k)=1/k$. Let $1>\varepsilon>0$ be a constant. 
    The preference probabilities are given by $\hat{p}_1 = (1, 1)^{\top} \in \mathbb{R}^{2\times 1}$, and $\hat{p}_2 = (1, 1-\varepsilon) \in \mathbb{R}^{1\times 2}$.
    This means that the single agent on $M$ equally prefers both agents on $N$, and that while both agents on $N$ strongly prefer the agent on $M$, $a_1$'s preference is slightly higher than $a_2$'s.
    
    For this instance, 
    %We show that a socially optimal policy may not double envy-free. Through a simple calculation, 
    we can see that policy $\pi_1 = \bigl(1,1,[[1,0],[0,1]]\bigr)$ is socially optimal.
    %where $1 \in \mathbb{R}^{1\times 1}$ and $I\in \mathbb{R}^{2\times 2}$. 
    Here, the components $1,1$ means that the single agent on $M$ is recommended to the two agents with probability $1$.
    The $[[1,0],[0,1]]$ component represents the recommendation matrix where $a_1$ is always recommended in the first position to the recommendation list of the agent on $M$, and $a_2$ is always recommended in the second position to the list.
    While this policy achieves the highest social welfare, it falls in terms of fairness. Specifically, it's not double envy-free since we have $U_2(C_2) = \frac{1-\varepsilon}{2} <  1-\varepsilon =U_2(C_1) $, and agent $a_2$ will envy agent $a_1$.
    
    On the other hand, policy $\pi_2 = \bigl(1,1,[[1/2,1/2],[1/2,1/2]]\bigr)$ satisfies double envy-freeness since we have $U_1(C_1) = U_1(C_1)$ and $U_2(C_2) = U_2(C_1)$. However, this policy is not socially optimal since $\mathrm{SW}(\pi_2) = \frac{3}{4} + \frac{3(1-\varepsilon)}{4} < 1+ \frac{1-\varepsilon}{2} = \mathrm{SW}(\pi_1)$.
\end{example}

\paragraph{Relationship to resource allocation problem}
It is worth noting that the problem of finding left-side envy-free opportunities in a special case is equivalent to the problem of finding an envy-free allocation in the fair division for divisible items.
Specifically, when $K=1$ and the matrices $A_1,A_2,\ldots,A_n$ are fixed, 
achieving left-side envy-freeness requires finding opportunities for all agents in $N$, represented as $C_1(1,\cdot),C_2(1,\cdot),\ldots,C_n(1,\cdot)$, such that
$($i$)$ for every pair of agents $a_{i},a_{i'}\in N$, 
$\sum_{j\in M}p_{ij}A_i(j,1) \cdot C_i(1,j) \geq \sum_{j\in M}p_{ij}A_i(j,1) \cdot C_{i'}(1,j)$,
$($ii$)$ $\sum_{i=1}^n C_i(1,j) = 1$ for each $j\in [m]$,
and $($iii$)$ $0\leq C_i(1,j) \leq 1$ for all $i\in [n], j\in [m]$.
In this case, $C_1(1,j),C_2(1,j),\ldots,C_n(1,j)$ indeed represent an envy-free allocation of left-side agents to right-side agents, where each agent $a_i\in N$ has a utility of $p_{ij}A_i(j,1)$ for each agent $b_j\in M$.

\section{Computing Socially Optimal Policy}\label{sec:sw}
In this section, we propose a reciprocal recommendation algorithm which compute a socially optimal policy approximately.

By utilizing $n^2m+nm^2$ variables $\{A_{i}(j,k)\}_{i,j,k}$ and $\{B_{j}(i,\ell)\}_{i,j,\ell}$,
our optimization problem of finding a socially optimal policy 
is to maximize 
$$
\mathrm{SW}(\bm{A},\bm{B}) = \sum_{i\in [n]} \sum_{j\in [m]} p_{i,j}  \sum_{k=1}^{m} \sum_{\ell=1}^{n} v_{k\ell} A_{i}(j,k) B_{j}(i,\ell)
$$ 
under the constraint that $A_i$ and $B_j$ are doubly stochastic matrices for all $i\in N$ and all $j\in M$.
Since the optimization objective is quadratic (non-linear), computing the global optimal solution is challenging.

Although the function $\mathrm{SW}(\bm{A},\bm{B})$ is not necessarily concave with respect to $(\bm{A},\bm{B})$, it is readily apparent that fixing one variable results in a concave function.
As a naive approach, one might naturally conceive of an alternating maximization method: first optimizing over $\bm{A}$ while keeping $\bm{B}$ fixed, and then optimizing over $\bm{B}$ with $\bm{A}$ fixed.
This procedure is essentially computing the best recommendations for each side alternately at each step. If the process converges to some values $(\bm{A}^*,\bm{B}^*)=(A_1^*,A_2^*,\ldots,A_n^*,B_1^*,B_2^*,\ldots,B_m^*)$, then $A_i^* = \mathrm{argmax}_{A_i}U_i (A_{i},\bm{B}^*)$ and 
$B_j^* = \mathrm{argmax}_{B_j}U_j (\bm{A}^*,B_{j})$
for all $i\in N$ and for all $j\in M$.
From linearity of the max function, we also have 
$\bm{A}^* \in \mathrm{argmax}_{\bm{A}}\mathrm{SW} (\bm{A},\bm{B}^*)$
and 
$\bm{B}^* \in \mathrm{argmax}_{\bm{B}}\mathrm{SW} (\bm{A}^*,\bm{B})$.

We now introduce a general algorithm for alternating optimization, as outlined in Algorithm~\ref{alg:alternating_SW_maximization}, which employs the Frank-Wolfe algorithm to solve the convex optimization problems~\cite{Frank1956,Su2022}. This framework will also be utilized in the subsequent section. 
We consider two functions $F_1$ and $F_2$, which are real-valued functions that take $\bm{A}$ and $\bm{B}$ as arguments. 
Moreover, we assume that the function $F_1(\bm{A},\bm{B})$ is concave with respect to $\bm{B}$, while the function $F_2(\bm{A},\bm{B})$ is concave with respect to $\bm{A}$. Thus, under the condition where $\bm{A}$ (resp. $\bm{B}$) is fixed, the function $F_1(\bm{A},\bm{B})$ (resp. $F_1(\bm{A},\bm{B})$) can be maximized with respect to $\bm{B}$ (resp. $\bm{A}$) using the Frank-Wolfe algorithm.

In the algorithm, the initialization of doubly stochastic matrices $\bm{A}$ and $\bm{B}$ is performed in line~\ref{line:initialize_A_SW}. 
In the \textbf{for} loop (lines~\ref{line:loop_start_A_SW}), the alternating optimization process is carried out, and the process continues until convergence is achieved or a predetermined number of iterations is reached.

To utilize this framework for the alternating social welfare maximization, the functions $F_1$ and $F_2$ are set as $F_1(\bm{A},\bm{B})=F_2(\bm{A},\bm{B})=\mathrm{SW}(\bm{A},\bm{B})$.
If the values of $\bm{A}$ and $\bm{B}$ converge in the alternating algorithm, we obtain a policy ensuring that all agents receive their best recommendations.
Although it is anticipated that this procedure may converge to a local optimum or fail to converge, experiments outlined in Section~\ref{sec:experiments} demonstrate convergence and the ability to achieve high social welfare values.
\begin{algorithm}[h]
    \caption{Alternating Maximization of Functions $F_1(\bm{A},\bm{B})$ and $F_2(\bm{A},\bm{B})$  via Frank-Wolfe Algorithm}
    \label{alg:alternating_SW_maximization}
    \begin{algorithmic}[1]
        \renewcommand{\algorithmicrequire}{\textbf{Input:}}
        \renewcommand{\algorithmicensure}{\textbf{Output:}}
        \REQUIRE preference probabilities $(p_{i,j})_{i,j}$, examination function $v(\cdot)$, learning rate $(\eta_t)_{t\in [T]}$.
        %\ENSURE doubly stochastic matrices $\bm{A},\bm{B}$.
        \STATE Initialize $\bm{A},\bm{B}$.\label{line:initialize_A_SW}
        \FOR{$t=1,2,\ldots,T$} \label{line:loop_start_A_SW}
        \STATE $\bm{X}^* \in \text{argmax}_{\bm{X}} \frac{\partial}{\partial \bm{A}} F_2(\bm{A},\bm{B})^{\top} \bm{X}$ s.t. $\bm{X}_{i}\boldsymbol{1} =\boldsymbol{1}$, $\boldsymbol{1}^{\top}\bm{X}_{i} = \boldsymbol{1}^{\top}$, $\bm{X}_{i} \in \mathbb{R}^{m\times m}_{\geq 0}$ $\forall i\in N$, and $\bm{X} = (\bm{X}_{i})_{i\in N}$.\label{line:maximize_A_SW}
        \STATE $\bm{A} \leftarrow (1-\eta_t)\bm{A}  + \eta_t \bm{X}^*$.
        \STATE $\bm{Y}^* \in \text{argmax}_{\bm{Y}} \frac{\partial}{\partial \bm{B}} F_1(\bm{A},\bm{B})^{\top} \bm{Y}$ s.t. $\bm{Y}_{j}\boldsymbol{1} =\boldsymbol{1}$, $\boldsymbol{1}^{\top}\bm{Y}_{j} = \boldsymbol{1}^{\top}$, $\bm{Y}_{j} \in \mathbb{R}^{n\times n}_{\geq 0}$ $\forall j\in M$, and $\bm{Y} = (\bm{Y}_{j})_{j\in M}$.\label{line:maximize_B_SW}
        \STATE $\bm{B} \leftarrow (1-\eta_t)\bm{B}  + \eta_t \bm{Y}^*$.
        %\IF{} break
        %\ENDIF
        \ENDFOR \label{line:while_end_SW}
        \RETURN $\bm{A},\bm{B}$.
    \end{algorithmic}
\end{algorithm}

\section{COMPUTING Fair Recommendation}\label{sec:fair_recommendation}
As illustrated in Example~\ref{example:non-envy-free}, socially optimal policies do not always guarantee double envy-freeness.
This observation is further supported by the experimental results in Section~\ref{sec:experiments}, which reveal that the policies derived from Algorithm~\ref{alg:alternating_SW_maximization} and other algorithms designed to maximize social welfare can result in significant levels of envy across both sides of the market.

We present a novel approach that aims to balance fairness and high social welfare in RRSs. 
We propose a method that aims to achieve a double envy-free policy by utilizing the \emph{Nash social welfare (NSW)} functions~\cite{Eisenberg1959,Varian1974,Kroer2019}. 
This approach is an extension of the work by Saito and Joachims~\cite{SaitoJoachimsKDD2022}, who considered the problem of recommending items to users in a one-sided recommendation setting. 
We adapt their technique to the reciprocal recommendation problem, where both sides of the market receive recommendations.

We introduce two NSW-style functions.
For policy $\pi=(\bm{A},\bm{B})$, the \emph{left-side NSW} function $\mathrm{NSW}_1(\bm{A}, \bm{B})$ is defined as the product of the expected utilities that each agent in $N$ receives under policy $\pi$. Formally,
\begin{align*}
    \mathrm{NSW}_1(\bm{A}, \bm{B}) 
    = \prod_{i\in [n]} U_i(\pi) 
    = \prod_{a\in N} \sum_{b\in M} \mathrm{Pr}[\text{$a$ matches with $b$} ]. 
    %&= \prod_{j\in M} \sum_{i\in N} p_{i,j}  \sum_{k=1}^{m} \sum_{\ell=1}^{n} v_{k\ell} A_{i}(j,k) B_{j}(i,\ell).
\end{align*}
Similarly, the \emph{right-side NSW} function $\mathrm{NSW}_2(\bm{A}, \bm{B})$ for policy $\pi=(\bm{A},\bm{B})$ is given by
\begin{align*}
    \mathrm{NSW}_2(\bm{A},\bm{B}) 
    = \prod_{j\in [m]} U_j(\pi) 
    = \prod_{b\in M} \sum_{a\in N}  \mathrm{Pr}[\text{$b$ matches with $a$} ].
    %&= \prod_{i\in N} \sum_{j\in M} p_{i,j}  \sum_{k=1}^{m} \sum_{\ell=1}^{n} v_{k\ell} A_{i}(j,k) B_{j}(i,\ell).
\end{align*}

Intuitively, since the left-side and right-side NSW functions consider the product of utilities, they naturally select the most balanced set of utilities. For instance, for left-side agents $N$, maximizing the left-side NSW function ($\mathrm{NSW}_1$) implies that the they are recommended to the right-side agents in a manner that they are guaranteed fairness and the higher welfare.
We say that for the left-side agents, the set $\bm{B}^*=(B_1^*,B_2^*,\ldots,B_m^*)$ of recommendations is \emph{NSW-best} if $\bm{B}^*$ maximizes $\mathrm{NSW}_1(\bm{A}, \bm{B})$ for given $\bm{A}$.
Similarly,
for the right-side agents, the set $\bm{A}^*=(A_1^*,A_2^*,\ldots,A_n^*)$ of recommendations is \emph{NSW-best} if $\bm{A}^*$ maximizes $\mathrm{NSW}_2 (\bm{A}, \bm{B})$.

Theorem 4.1 of Saito and Joachims~\cite{SaitoJoachimsKDD2022} and Theorem 1 of Kroer and Peysakhovich~\cite{Kroer2019} immediately imply that if two sets of recommendations simultaneously satisfy NSW-best under a certain condition, then the policy is approximately double envy-free.

To elucidate this, we introduce a definition pertaining to the preferences of agents. 
The values of preference probabilities $(p_{i,j})_{i,j}$ satisfies $\varepsilon$-\emph{similarity} for a nonnegative constant $\varepsilon\geq 0$ if for every left-side agent $a_i\in N$, there exists a set of the left-side agents $N'\subseteq N$ of size at least $K+1$ where each left-side agent $a_{i'}\in N'$ satisfies $\max_{j\in [m]}|p_{i,j} - p_{i',j}|\leq \varepsilon$, and, simultaneously, for every right-side agent $b_j\in M$, there exists a subset $M'\subseteq M$ of size at least $K+1$ where each right-side agent $b_{j'}\in M'$ satisfies $\max_{i\in [n]}|p_{i,j} - p_{i,j'}|\leq \varepsilon$.
When $n$ and $m$ are substantially larger than $K$, the assumption that there exist at most a constant number $K+1$ of similar individuals for each agent in the market does not impose an excessively strict constraint.
Then, we present the following theorem.
\begin{theorem}\label{thm:envy-free-policy}
    For the case of $K=1$, if a policy induces NSW-best recommendations for both sides, then the policy is double envy-free.
    %
    %Let $\varepsilon\geq 0$ be a constant.
    For the case of $K>1$, letting $\varepsilon$ be a non-negative constant, we assume that the estimated preference probabilities satisfies $\varepsilon$-similarity, and $n=\Theta(m)$\footnote{I.e., there exist two positive constants $c_1$ and $c_2$ such that $c_1\cdot m \leq n \leq c_2\cdot m$.}. If a policy $\pi=(\bm{A},\bm{B})$ induces NSW-best recommendations for both sides,
    then the policy satisfies double envy-freeness up to an additive difference of $\mathrm{O}(\varepsilon)$, that is, for every pair of the left-side agents $a_{i},a_{i'}\in N$,
        $U_i\bigl(C_i\bigr) \geq U_i\bigl(C_{i'}\bigr) - \mathrm{O}(\varepsilon)$,
    and
    for every pair of the right-side agents $b_{j},b_{j'}\in M$,
        $U_j\bigl(D_j\bigr)\geq  U_j\bigl(D_{j'}\bigr) - \mathrm{O}(\varepsilon)$. 
\end{theorem}

We now discuss a method for finding Nash Social Welfare (NSW)-best recommendations for both sides. We employ the alternating optimization technique outlined in Algorithm~\ref{alg:alternating_SW_maximization}.

When $\bm{A}$ is fixed, $\log \mathrm{NSW}_2(\bm{A},\bm{B})$ is a concave function over $\bm{B}$, as it is a sum of logarithms of sums. Similarly, when $\bm{B}$ is fixed, $\log \mathrm{NSW}_1(\bm{A},\bm{B})$ is a concave function over $\bm{A}$.
Therefore, we can apply the alternating maximization algorithm
by setting $F_1(\bm{A},\bm{B}) = \log \mathrm{NSW}_1(\bm{A},\bm{B})$ and $F_2(\bm{A},\bm{B}) = \log \mathrm{NSW}_2(\bm{A},\bm{B})$ in Algorithm~\ref{alg:alternating_SW_maximization}.
In the alternating maximization algorithm,
if $(\bm{A},\bm{B})$ converges to some value
$(\bm{A}^*,\bm{B}^*)$, then $\bm{A}^*$ and  $\bm{B}^*$ are both NSW-best. From Theorem~\ref{thm:envy-free-policy}, we obtain a recommendation policy satisfying double envy-freeness up to an additive difference of $\mathrm{O}(\varepsilon)$.
The experimental results presented in Section~\ref{sec:experiments} provide strong empirical evidence for the effectiveness of our approach. For detailed results, refer to Sections~\ref{sec:experiments:Synthetic_Data:results} and~\ref{sec:experiments:Real_World_Data:results}.

\paragraph{A heuristic method based on the matching problem}We note the connection between matchings on a bipartite graph and reciprocal recommendations for both sides, and propose a method based on maximum weight matchings.
In fact, a matching on the complete bipartite graph between $N$ and $M$ with edge-weights $(p_{i,j})_{i,j}$ provides recommendations. 
Here, for some $k \in [K]$, $A_{i}(j,k) = 1$ and $B_{j}(i,k) = 1$ if and only if the edge $\{i,j\}$ is included in the matching.
It is north noting that, in the special case where $K = 1$, $n = m$ and only deterministic policies are considered, a policy induced from a maximum weight matching is socially optimal among deterministic policies. Furthermore, the policy satisfies double envy-freeness since if agent $a_i\in N$ matches with agent $b_j\in M$, then we have $B_j(i',1)=0$ for $i'\neq i$ and another agent $a_{i'} \in N$ never envies agent $a_{i}$. A similar discussion holds for agents in $M$. 

We propose an algorithm based on maximum weight matchings (Algorithm~\ref{alg:IterLP}) that iteratively solves a linear programming (LP) relaxation problem for the maximum weight matching, filling up each position $k$ of recommendations. At each iteration, the algorithm recommends the most preferred partner for each agent based on the solution of the LP.

\begin{algorithm}[h]
    \caption{Iterating Linear Programming of Maximum Weight Matching}
    \label{alg:IterLP}
    \begin{algorithmic}[1]
        \renewcommand{\algorithmicrequire}{\textbf{Input:}}
        \renewcommand{\algorithmicensure}{\textbf{Output:}}
        \REQUIRE: preference probabilities $(p_{i,j})_{i,j}$.
        %\ENSURE: doubly stochastic matrices $\bm{A},\bm{B}$.
        \STATE Initialize $A_i=O,B_j=O$ for all $i\in N,j\in M$.
        \FOR{$k=1,2,\ldots,K$} 
        \STATE Let $(x_{i,j}^*)_{i,j}$ be an LP solution of the maximum weight matching between $N$ and $M$ with edge-weights $(p_{i,j})_{i,j}$.
        \STATE For each $i\in N$, let $j^*_i = \max_{j\in M} x_{i,j}^*$, and set $A_i(j^*_i,k)=1$.
        \STATE For each $j\in M$, let $i^*_j = \max_{i\in N} x_{i,j}^*$, and set $B_j(i^*_j ,k)=1$.
        \STATE Set the weights of $\{i,j^*_i\}$ and $\{i^*_j,j\}$ to $-\infty$.
        \ENDFOR \label{line:for_end}
        \RETURN $\bm{A},\bm{B}$.
    \end{algorithmic}
\end{algorithm}

\section{Experiments}\label{sec:experiments}

In this section, we report the experimental results with synthetic data and real-world data from a Japanese online dating platform\footnote{The codes, implemented in Python, are available at \url{https://github.com/CyberAgentAILab/FairReciprocalRecommendation}}.

\subsection{Synthetic Data}

\begin{figure*}[t]
    \begin{center}
    \includegraphics[width=\linewidth]{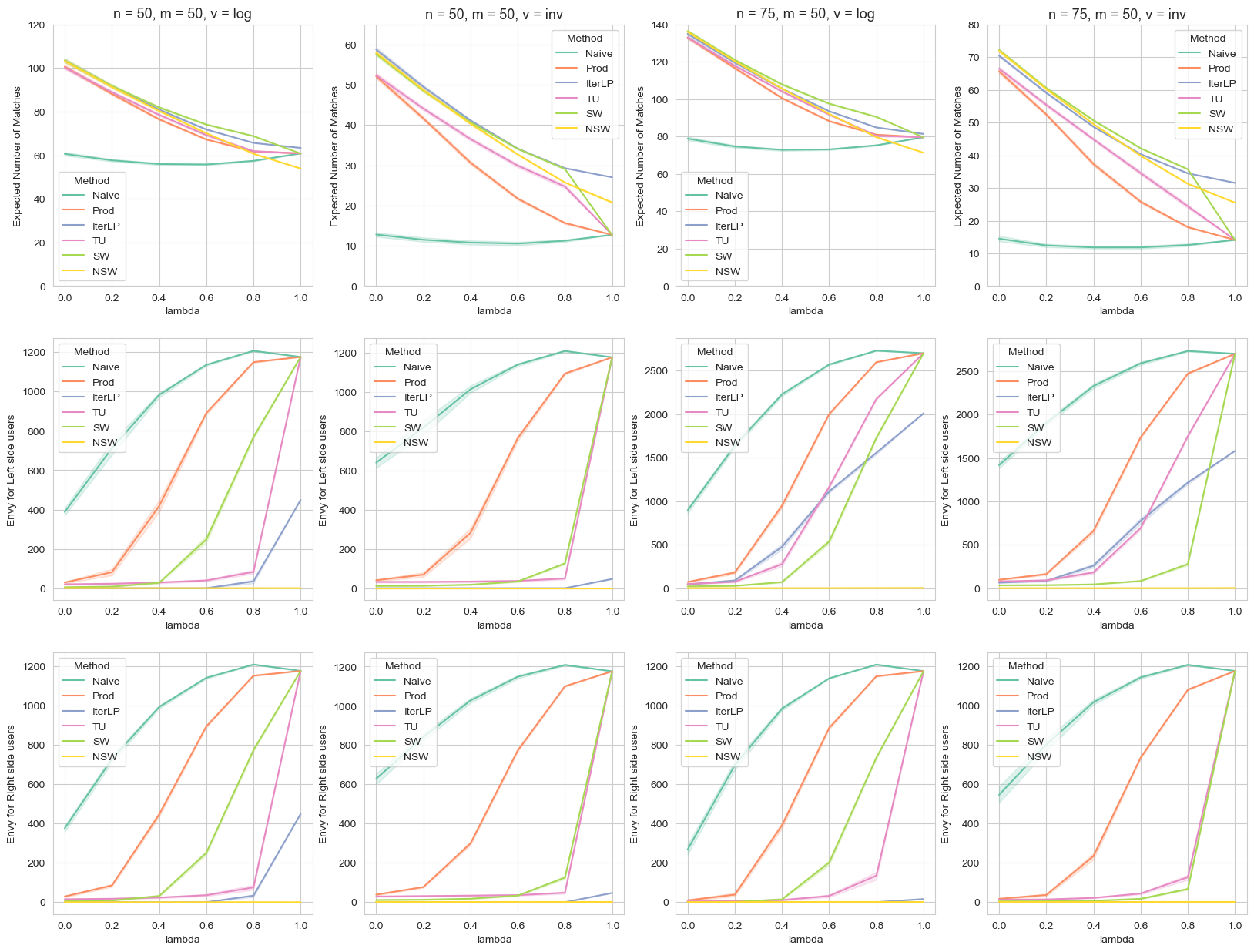}
    \caption{Results of experiments on synthetic data.
    Upper, middle and lower rows show the expected number of matches, the number of envies for left-side agents, and that for right-side agents for each case, respectively.
    We varies the number of left-side agents $n \in \{50, 75\}$, the examination functions $v = 1/\log_2(k+1)$ (log) or $1/k$ (inv), and lambda $\lambda \in \{0.0, 0.2, \dots, 1.0\}$, while we fixed the number of right-side agents $m = 50$, where each case correspond to each column.
    We conducted 10 experiments for each case, and report the mean of results and its 95\% confidence intervals (but it is invisible in many cases due to its small variations.).}
    \label{fig:synthetic_data}
    \end{center}
\end{figure*}

\begin{figure*}[t]
    \begin{center}
    \includegraphics[width=0.8\linewidth]{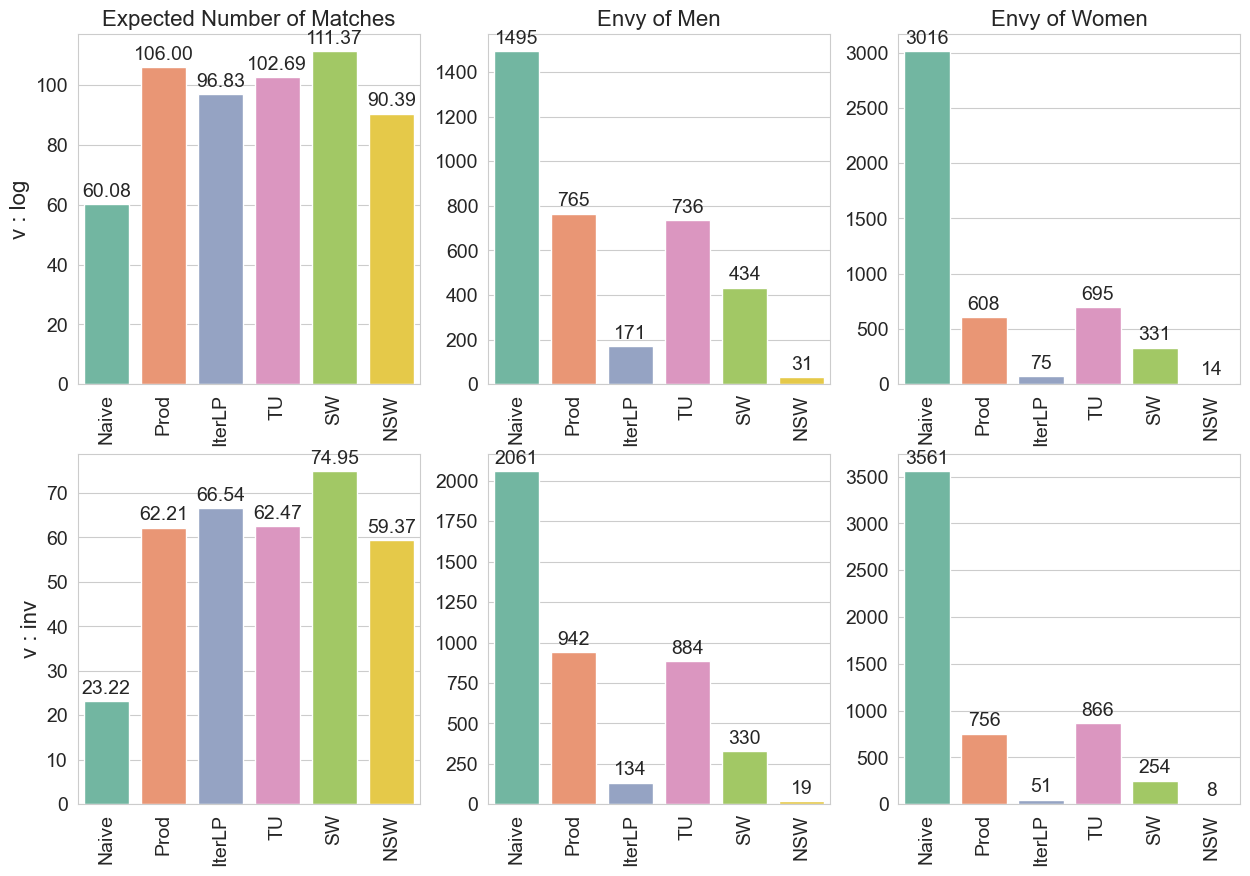}
    \caption{Results of experiments on the real-world data of a Japanese online dating platform. We conducted experiments $v(k) = 1/\log_2(k+1)$ (log) or $1/k$ (inv), and report the expected number of matches, the number of envy for male users and that of envy for female users.}
    \label{fig:realdata}
    \end{center}
\end{figure*}

\subsubsection{Data generation}
We first explain the generation procedure of synthetic data which represents an online dating platform data.
We set the number of right-side agents as $m = 50$, while we tests the case of the number of left-side agents as $n = 50$ (the balanced case) and that of $n = 75$ (the unbalanced case).
To control the level of differences of popularity among agents, we generate preference probabilities with two terms and a parameter $\lambda$ as \citet{su2022optimizing} and~\citet{Tomita2023} conducted:
$$
    \hat{p}_1(i,j) = \lambda \cdot \hat{p}_{1}^{\mathrm{pop}}(j) + (1-\lambda)\cdot \hat{p}_{1}^{\mathrm{unif}}(i,j)
$$
and 
$$
    \hat{p}_2(j,i) = \lambda \cdot \hat{p}_{2}^{\mathrm{pop}}(i) + (1-\lambda)\cdot \hat{p}_{2}^{\mathrm{unif}}(j,i)
$$
where $\hat{p}_{1}^{\mathrm{pop}}(j) = \frac{j-1}{m-1}$ and $\hat{p}_{2}^{\mathrm{pop}}(i) = \frac{i-1}{n-1}$ represent agent $b_j$'s overall popularity and agent $a_i$'s overall popularity respectively. Moreover, $\hat{p}_{1}^{\mathrm{unif}}(i,j)$ and $\hat{p}_{2}^{\mathrm{unif}}(j,i)$ represent individual preferences which are identically and independently drawn from the uniform distribution over $[0,1]$. 
The parameter $\lambda$ controls the level of differences in popularity, and we vary $\lambda \in \{0.0, 0.2, 0.4, 0.6, 0.8, 1.0\}$. 
When $\lambda$ is large, agents on each side of the market tend to have very similar preferences towards agents on the opposite side. 
This similarity in preferences leads to substantial differences in popularity among agents.
Finally, we tests two kinds of examination functions, $v(k) = \frac{1}{\log_2(k+1)}$ (``log'') or $v(k) = \frac{1}{k}$ (``inv'').

\subsubsection{Comparing methods}
We tests the following recommendation policies:
\begin{itemize}
    \item Naive: Recommend right-side agents to each left-side agent $a_i$ in the non-increasing order of the preference $\hat{p}_1(i, \cdot)$ deterministically, and similarly to right-side agents.
    \item Prod: Compute the reciprocal score of the pair $(i, j)$ by product $p_{i,j} = \hat{p}_1(i, j) \cdot \hat{p}_2(j, i)$, and recommend right-side agents to each left-side agent $a_i$ in the non-increasing order of this scores $p_{i, \cdot}$, and similarly to right-side agents. This method is widely used in practice as well as the Naive method~\cite{neve2019latent,neve2019aggregation}.
    \item IterLP: A heuristic method iterating Linear Programming (Algorithm~\ref{alg:IterLP}).
    \item TU: A deterministic policy based on TU matching model proposed by \citet{Tomita2023}.
    \item SW: Alternating social welfare maximization (Algorithm~\ref{alg:alternating_SW_maximization} with $F_1(\bm{A},\bm{B})=F_2(\bm{A},\bm{B})=\mathrm{SW}(\bm{A},\bm{B})$).
    \item NSW: Alternating Nash social welfare maximization (Algorithm~\ref{alg:alternating_SW_maximization} with $F_1(\bm{A},\bm{B})=\log \mathrm{NSW}_1(\bm{A},\bm{B})$ and $F_2(\bm{A},\bm{B})=\log \mathrm{NSW}_2(\bm{A},\bm{B})$).
\end{itemize}

\subsubsection{Experimental Procedure}
For each case of $n$ (the number of left-side agents), $\lambda$ (difference level of popularity) and examination functions, we generate 10 test samples and compute the expected number of matches, the number of pairs $(i, i')$ of two left-side agents such that $a_i$ has an envy towards $a_{i'}$, and the number of pairs $(j, j')$ of two right-side agents such that $b_j$ has an envy towards $b_{j'}$.
We report average numbers of 10 generated samples.

\subsubsection{Results}\label{sec:experiments:Synthetic_Data:results}

The results of experiments with synthetic data are summarized in Figure~\ref{fig:synthetic_data}.

We first note that as the popularity difference level $\lambda$ increases, the expected matches decrease and the number of envies increase for most methods.
This is because the recommendation problem becomes difficult as the difference level of popularity becomes high and the congestion of users occurred.
SW achieves the highest expected matches almost all cases unless $\lambda = 1.0$, while it occurs relatively many envies in case of $\lambda \ge 0.6$.
Thus we empirically show that there is trade-off between the social welfare and fairness of recommendation opportunities if there is some level of popularity differences among users.

Second, we emphasize that the NSW method achieves almost zero envies for all cases, while it achieves competitive level of expected matches with SW or others for most cases. NSW achieves high social welfare and zero envies even with the case of $\lambda = 0.8$ in which other methods occur high level of unfairness.

Finally, we note that the IterLP, which is a heuristic introduced in Algorithm~\ref{alg:IterLP},
%the appendix~\ref{sec:appendix}, 
achieves relatively high social welfare and small envies in many cases. However, it could not avoid envies for left-side agents in unbalanced case $n = 75$, where NSW achieves almost zero envies for both sides.

\subsection{Real-World Data}

To validate performances of the methods in practical situations, we also report the results of experiments with real-world data.

\subsubsection{A Japanese online dating platform data}

We conducted experiments using data from a Japanese online dating platform with millions of cumulative members. 
In this service, male users receive recommendation lists of female users and choose to send a ``like'', or not (``dislike'') to each user in the recommended list.
If a male user sends a ``like'' to a female user, she receives a notification and can choose to ``match'' with him or decline (``sorry'').
After a match is made, users are able to chat and communicate.
The process is similar for female users recommending and responding to male users.

To collect a moderate-sized dataset for our experiments and methods, we extracted data consisting of 200 male and 200 female users.
To alleviate the sparsity issue in training our Matrix Factorization-based preference model, we sampled a 200 $\times$ 200 user matrix where users had relatively more interactive actions with each other.
Using ``like''/``dislike'' data of male users and ``match''/``sorry'' data of female users in this sampled data, we compute preference scores $\hat{p}_1(i, j)$ and $\hat{p}_2(j, i)$ by the Alternating Least Squares (ALS) method~\cite{koren2009matrix}.

\subsubsection{Experiments and Results}\label{sec:experiments:Real_World_Data:results}

Using the preferences explained above, we conducted simulation experiments.
We compute the expected number of matches, the number of envies among male users and the number of envies among female users for each method.
We set the examination functions $v(k) = 1/\log_2(k+1)$ (``log'') or $1/k$ (``inv''), and comparing methods are similar to the case of synthetic data.

The results are shown in Figure~\ref{fig:realdata}. Similarly to the synthetic data experiments, SW achieves the highest social welfare (the expected number of matches), which NSW achieves competitive level of social welfare compared to Prod, IterLP and TU.
On the other hand, NSW occur clearly the smallest numbers of envies both for male users and female users.
IterLP also relatively small numbers of envies, but we note that the number of male users and female users are balanced to 200 $\times$ 200 in our sampled data and IterLP occur non-negligible envies for male users in unbalanced synthetic data.
Therefore we can conclude that the NSW method would be effective to alleviate unfairness of recommendation opportunities among users in practical online dating platforms.

\section{discussion}

In this study, we begin by presenting a model of reciprocal recommendations inspired by online dating platforms, and formalize the concepts of envy and fairness called envy-freeness.
We then define the platform's social welfare as the expected number of matches resulting from given recommendation policies, and explain the process of alternating social welfare (SW) maximization.
Our experiments reveals that while this SW maximization approach is effective in increasing matches, it can lead to unfairness in terms of envy-freeness. 
To address this, we propose the alternating Nash Social Welfare (NSW) maximization algorithm via the Frank-Wolfe algorithm.

Through experiments with both synthetic and real-world data, we demonstrate that the NSW maximization method results in almost zero instances of envy while achieving relatively high levels of social welfare compared to other existing methods.
These results suggest that the NSW method would be effective in two-sided matching platforms where fairness among users is crucial such as online dating platforms.

However, our experiments also revealed scalability challenges.
While heuristics like IterLP and TU operate efficiently for large problems, the alternating NSW maximization via Frank-Wolfe requires handling $n^2m+nm^2$ variables, causing substantial computational costs as $n$ and $m$ increase. This potentially limits the NSW method's practical application in large-scale scenarios. 
Future work should focus on developing methods to compute fair reciprocal recommendations efficiently for larger models.

\bibliographystyle{ACM-Reference-Format}
\bibliography{main}


\begin{thebibliography}{52}

%%% the Web address.

\ifx \showCODEN    \undefined \def \showCODEN     #1{\unskip}     \fi
\ifx \showDOI      \undefined \def \showDOI       #1{#1}\fi
\ifx \showISBNx    \undefined \def \showISBNx     #1{\unskip}     \fi
\ifx \showISBNxiii \undefined \def \showISBNxiii  #1{\unskip}     \fi
\ifx \showISSN     \undefined \def \showISSN      #1{\unskip}     \fi
\ifx \showLCCN     \undefined \def \showLCCN      #1{\unskip}     \fi
\ifx \shownote     \undefined \def \shownote      #1{#1}          \fi
\ifx \showarticletitle \undefined \def \showarticletitle #1{#1}   \fi
\ifx \showURL      \undefined \def \showURL       {\relax}        \fi
% invisible to TeX
\providecommand\bibfield[2]{#2}
\providecommand\bibinfo[2]{#2}
\providecommand\natexlab[1]{#1}
\providecommand\showeprint[2][]{arXiv:#2}

\bibitem[Balcan et~al\mbox{.}(2019)]%
        {balcan2019envyfree}
\bibfield{author}{\bibinfo{person}{Maria-Florina~F Balcan}, \bibinfo{person}{Travis Dick}, \bibinfo{person}{Ritesh Noothigattu}, {and} \bibinfo{person}{Ariel~D Procaccia}.} \bibinfo{year}{2019}\natexlab{}.
\newblock \showarticletitle{Envy-free classification}. In \bibinfo{booktitle}{\emph{Advances in Neural Information Processing Systems}}, Vol.~\bibinfo{volume}{32}. \bibinfo{pages}{1240--1250}.
\newblock


\bibitem[Beutel et~al\mbox{.}(2019)]%
        {beutel2019fairness}
\bibfield{author}{\bibinfo{person}{Alex Beutel}, \bibinfo{person}{Jilin Chen}, \bibinfo{person}{Tulsee Doshi}, \bibinfo{person}{Hai Qian}, \bibinfo{person}{Li Wei}, \bibinfo{person}{Yi Wu}, \bibinfo{person}{Lukasz Heldt}, \bibinfo{person}{Zhe Zhao}, \bibinfo{person}{Lichan Hong}, \bibinfo{person}{Ed~H. Chi}, {and} \bibinfo{person}{Cristos Goodrow}.} \bibinfo{year}{2019}\natexlab{}.
\newblock \showarticletitle{Fairness in recommendation ranking through pairwise comparisons}. In \bibinfo{booktitle}{\emph{Proceedings of the 25th ACM SIGKDD International Conference on Knowledge Discovery and Data Mining}}. \bibinfo{pages}{2212--2220}.
\newblock


\bibitem[Biega et~al\mbox{.}(2018)]%
        {biega2018equity}
\bibfield{author}{\bibinfo{person}{Asia~J Biega}, \bibinfo{person}{Krishna~P Gummadi}, {and} \bibinfo{person}{Gerhard Weikum}.} \bibinfo{year}{2018}\natexlab{}.
\newblock \showarticletitle{Equity of Attention: Amortizing Individual Fairness in Rankings}. In \bibinfo{booktitle}{\emph{The 41st International ACM SIGIR Conference on Research \& Development in Information Retrieval}}. \bibinfo{pages}{405--414}.
\newblock


\bibitem[Budish(2011)]%
        {Budish2011}
\bibfield{author}{\bibinfo{person}{Eric Budish}.} \bibinfo{year}{2011}\natexlab{}.
\newblock \showarticletitle{The combinatorial assignment problem: approximate competitive equilibrium from equal incomes}.
\newblock \bibinfo{journal}{\emph{Journal of Political Economy}} \bibinfo{volume}{119}, \bibinfo{number}{6} (\bibinfo{year}{2011}), \bibinfo{pages}{1061--1103}.
\newblock


\bibitem[Caragiannis et~al\mbox{.}(2019)]%
        {Caragiannis2019}
\bibfield{author}{\bibinfo{person}{Ioannis Caragiannis}, \bibinfo{person}{David Kurokawa}, \bibinfo{person}{Herv\'{e} Moulin}, \bibinfo{person}{Ariel~D. Procaccia}, \bibinfo{person}{Nisarg Shah}, {and} \bibinfo{person}{Junxing Wang}.} \bibinfo{year}{2019}\natexlab{}.
\newblock \showarticletitle{{The unreasonable fairness of maximum Nash welfare}}.
\newblock \bibinfo{journal}{\emph{ACM Transactions on Economics and Computation}} \bibinfo{volume}{7}, \bibinfo{number}{3} (\bibinfo{year}{2019}), \bibinfo{pages}{1--12}.
\newblock


\bibitem[Celdir et~al\mbox{.}(2024)]%
        {celdir2024popularity}
\bibfield{author}{\bibinfo{person}{Musa~Eren Celdir}, \bibinfo{person}{Soo-Haeng Cho}, {and} \bibinfo{person}{Elina~H Hwang}.} \bibinfo{year}{2024}\natexlab{}.
\newblock \showarticletitle{Popularity bias in online dating platforms: Theory and empirical evidence}.
\newblock \bibinfo{journal}{\emph{Manufacturing \& Service Operations Management}} \bibinfo{volume}{26}, \bibinfo{number}{2} (\bibinfo{year}{2024}), \bibinfo{pages}{537--553}.
\newblock


\bibitem[Chen et~al\mbox{.}(2023)]%
        {chen2023reducing}
\bibfield{author}{\bibinfo{person}{Kuan-Ming Chen}, \bibinfo{person}{Yu-Wei Hsieh}, {and} \bibinfo{person}{Ming-Jen Lin}.} \bibinfo{year}{2023}\natexlab{}.
\newblock \showarticletitle{Reducing recommendation inequality via two-sided matching: A field experiment of online dating}.
\newblock \bibinfo{journal}{\emph{International Economic Review}} \bibinfo{volume}{64}, \bibinfo{number}{3} (\bibinfo{year}{2023}), \bibinfo{pages}{1201--1221}.
\newblock


\bibitem[Cole and Gkatzelis(2018)]%
        {ColeGkatzelis2018}
\bibfield{author}{\bibinfo{person}{Richard Cole} {and} \bibinfo{person}{Vasilis Gkatzelis}.} \bibinfo{year}{2018}\natexlab{}.
\newblock \showarticletitle{Approximating the {Nash} social welfare with indivisible items}.
\newblock \bibinfo{journal}{\emph{SIAM J. Comput.}} \bibinfo{volume}{47}, \bibinfo{number}{3} (\bibinfo{year}{2018}), \bibinfo{pages}{1211--1236}.
\newblock


\bibitem[Craswell et~al\mbox{.}(2008)]%
        {craswell2008experimental}
\bibfield{author}{\bibinfo{person}{Nick Craswell}, \bibinfo{person}{Onno Zoeter}, \bibinfo{person}{Michael Taylor}, {and} \bibinfo{person}{Bill Ramsey}.} \bibinfo{year}{2008}\natexlab{}.
\newblock \showarticletitle{An Experimental Comparison of Click Position-bias Models}. In \bibinfo{booktitle}{\emph{Proceedings of the 2008 International Conference on Web Search and Data Mining}}. \bibinfo{pages}{87--94}.
\newblock


\bibitem[Deldjoo et~al\mbox{.}(2024)]%
        {DeldjooJannach2024}
\bibfield{author}{\bibinfo{person}{Yashar Deldjoo}, \bibinfo{person}{Dietmar Jannach}, \bibinfo{person}{Alejandro Bellog^^c3^^adn}, \bibinfo{person}{Alessandro Difonzo}, {and} \bibinfo{person}{Dario Zanzonelli}.} \bibinfo{year}{2024}\natexlab{}.
\newblock \showarticletitle{Fairness in recommender systems: research landscape and future directions}.
\newblock \bibinfo{journal}{\emph{User Modeling and User-Adapted Interaction}}  \bibinfo{volume}{34} (\bibinfo{year}{2024}), \bibinfo{pages}{59--108}.
\newblock


\bibitem[Do et~al\mbox{.}(2021)]%
        {Virginie2021}
\bibfield{author}{\bibinfo{person}{Virginie Do}, \bibinfo{person}{Sam Corbett-Davies}, \bibinfo{person}{Jamal Atif}, {and} \bibinfo{person}{Nicolas Usunier}.} \bibinfo{year}{2021}\natexlab{}.
\newblock \showarticletitle{Two-sided fairness in rankings via Lorenz dominance}. In \bibinfo{booktitle}{\emph{Advances in Neural Information Processing Systems}}, Vol.~\bibinfo{volume}{34}. \bibinfo{publisher}{Curran Associates, Inc.}, \bibinfo{pages}{8596--8608}.
\newblock


\bibitem[Do et~al\mbox{.}(2022)]%
        {do2022online}
\bibfield{author}{\bibinfo{person}{Virginie Do}, \bibinfo{person}{Sam Corbett-Davies}, \bibinfo{person}{Jamal Atif}, {and} \bibinfo{person}{Nicolas Usunier}.} \bibinfo{year}{2022}\natexlab{}.
\newblock \showarticletitle{Online certification of preference-based fairness for personalized recommender systems}. In \bibinfo{booktitle}{\emph{AAAI Conference on Artificial Intelligence}}, Vol.~\bibinfo{volume}{36}. \bibinfo{pages}{6532--6540}.
\newblock


\bibitem[Eisenberg and Gale(1959)]%
        {Eisenberg1959}
\bibfield{author}{\bibinfo{person}{Edmund Eisenberg} {and} \bibinfo{person}{David Gale}.} \bibinfo{year}{1959}\natexlab{}.
\newblock \showarticletitle{Consensus of subjective probabilities: the Pari-mutuel method}.
\newblock \bibinfo{journal}{\emph{The Annals of Mathematical Statistics}} \bibinfo{volume}{30}, \bibinfo{number}{1} (\bibinfo{year}{1959}), \bibinfo{pages}{165--168}.
\newblock


\bibitem[Ekstrand et~al\mbox{.}(2018)]%
        {ekstrand2018all}
\bibfield{author}{\bibinfo{person}{Michael~D. Ekstrand}, \bibinfo{person}{Mucun Tian}, \bibinfo{person}{Ion~Madrazo Azpiazu}, \bibinfo{person}{Jennifer~D. Ekstrand}, \bibinfo{person}{Oghenemaro Anuyah}, \bibinfo{person}{David McNeill}, {and} \bibinfo{person}{Maria~Soledad Pera}.} \bibinfo{year}{2018}\natexlab{}.
\newblock \showarticletitle{All The Cool Kids, How Do They Fit In?: Popularity and Demographic Biases in Recommender Evaluation and Effectiveness}. In \bibinfo{booktitle}{\emph{Proceedings of the 1st Conference on Fairness, Accountability and Transparency}}, Vol.~\bibinfo{volume}{81}. \bibinfo{pages}{172--186}.
\newblock


\bibitem[Foley(1967)]%
        {Foley1967}
\bibfield{author}{\bibinfo{person}{Duncan~K. Foley}.} \bibinfo{year}{1967}\natexlab{}.
\newblock \showarticletitle{{Resource allocation and the public sector}}.
\newblock \bibinfo{journal}{\emph{Yale Economic Essays}}  \bibinfo{volume}{7} (\bibinfo{year}{1967}), \bibinfo{pages}{45--98}.
\newblock


\bibitem[Frank and Wolfe(1956)]%
        {Frank1956}
\bibfield{author}{\bibinfo{person}{Marguerite Frank} {and} \bibinfo{person}{Philip Wolfe}.} \bibinfo{year}{1956}\natexlab{}.
\newblock \showarticletitle{An algorithm for quadratic programming}.
\newblock \bibinfo{journal}{\emph{Naval Research Logistics Quarterly}}  \bibinfo{volume}{3} (\bibinfo{year}{1956}), \bibinfo{pages}{95}.
\newblock


\bibitem[Freeman et~al\mbox{.}(2021)]%
        {freeman2021two}
\bibfield{author}{\bibinfo{person}{Rupert Freeman}, \bibinfo{person}{Evi Micha}, {and} \bibinfo{person}{Nisarg Shah}.} \bibinfo{year}{2021}\natexlab{}.
\newblock \showarticletitle{Two-sided matching meets fair division}. In \bibinfo{booktitle}{\emph{Proceedings of the 30th International Joint Conference on Artificial Intelligence}}. \bibinfo{pages}{203--209}.
\newblock


\bibitem[Garg et~al\mbox{.}(2023)]%
        {Garg2023}
\bibfield{author}{\bibinfo{person}{Jugal Garg}, \bibinfo{person}{Edin Husi\'{c}}, \bibinfo{person}{Wenzheng Li}, \bibinfo{person}{L\'{a}szl\'{o}~A V\'{e}gh}, {and} \bibinfo{person}{Jan Vondr\'{a}k}.} \bibinfo{year}{2023}\natexlab{}.
\newblock \showarticletitle{Approximating Nash social welfare by matching and local search}. In \bibinfo{booktitle}{\emph{Proceedings of the 55th {{Annual ACM Symposium}} on {{Theory}} of {{Computing}}}}. \bibinfo{pages}{1298--1310}.
\newblock


\bibitem[Gollapudi et~al\mbox{.}(2020)]%
        {gollapudi2020almost}
\bibfield{author}{\bibinfo{person}{Sreenivas Gollapudi}, \bibinfo{person}{Kostas Kollias}, {and} \bibinfo{person}{Benjamin Plaut}.} \bibinfo{year}{2020}\natexlab{}.
\newblock \showarticletitle{Almost envy-free repeated matching in two-sided markets}. In \bibinfo{booktitle}{\emph{Proceedings of the 16th International Conference on Web and Internet Economics}}. \bibinfo{pages}{3--16}.
\newblock


\bibitem[G\"{o}lz et~al\mbox{.}(2019)]%
        {golz2019paradoxes}
\bibfield{author}{\bibinfo{person}{Paul G\"{o}lz}, \bibinfo{person}{Anson Kahng}, {and} \bibinfo{person}{Ariel~D Procaccia}.} \bibinfo{year}{2019}\natexlab{}.
\newblock \showarticletitle{Paradoxes in fair machine learning}. In \bibinfo{booktitle}{\emph{Advances in Neural Information Processing Systems}}, Vol.~\bibinfo{volume}{32}.
\newblock


\bibitem[Igarashi et~al\mbox{.}(2023)]%
        {igarashi2023fair}
\bibfield{author}{\bibinfo{person}{Ayumi Igarashi}, \bibinfo{person}{Yasushi Kawase}, \bibinfo{person}{Warut Suksompong}, {and} \bibinfo{person}{Hanna Sumita}.} \bibinfo{year}{2023}\natexlab{}.
\newblock \showarticletitle{Fair division with two-sided preferences}. In \bibinfo{booktitle}{\emph{Proceedings of the 32nd International Joint Conference on Artificial Intelligence}}. \bibinfo{pages}{2756--2764}.
\newblock


\bibitem[Jain and Vaish(2024)]%
        {JainVaish2024}
\bibfield{author}{\bibinfo{person}{Pallavi Jain} {and} \bibinfo{person}{Rohit Vaish}.} \bibinfo{year}{2024}\natexlab{}.
\newblock \showarticletitle{Maximizing {Nash} social welfare under two-sided preferences}. In \bibinfo{booktitle}{\emph{Proceedings of the 38th {AAAI} Conference on Artificial Intelligence}}.
\newblock


\bibitem[J{\"a}rvelin and Kek{\"a}l{\"a}inen(2002)]%
        {jarvelin2002cumulated}
\bibfield{author}{\bibinfo{person}{Kalervo J{\"a}rvelin} {and} \bibinfo{person}{Jaana Kek{\"a}l{\"a}inen}.} \bibinfo{year}{2002}\natexlab{}.
\newblock \showarticletitle{Cumulated Gain-based Evaluation of IR Techniques}.
\newblock \bibinfo{journal}{\emph{ACM Transactions on Information Systems}} \bibinfo{volume}{20}, \bibinfo{number}{4} (\bibinfo{year}{2002}), \bibinfo{pages}{422--446}.
\newblock


\bibitem[Joachims et~al\mbox{.}(2017)]%
        {Joachims2017-ix}
\bibfield{author}{\bibinfo{person}{Thorsten Joachims}, \bibinfo{person}{Adith Swaminathan}, {and} \bibinfo{person}{Tobias Schnabel}.} \bibinfo{year}{2017}\natexlab{}.
\newblock \showarticletitle{Unbiased learning-to-rank with biased feedback}. In \bibinfo{booktitle}{\emph{Proceedings of the tenth ACM international conference on web search and data mining}}. \bibinfo{pages}{781--789}.
\newblock


\bibitem[Kearns et~al\mbox{.}(2018)]%
        {kearns2018preventing}
\bibfield{author}{\bibinfo{person}{Michael Kearns}, \bibinfo{person}{Seth Neel}, \bibinfo{person}{Aaron Roth}, {and} \bibinfo{person}{Zhiwei~Steven Wu}.} \bibinfo{year}{2018}\natexlab{}.
\newblock \showarticletitle{Preventing fairness gerrymandering: auditing and learning for subgroup fairness}. In \bibinfo{booktitle}{\emph{Proceedings of the 35th International Conference on Machine Learning}}. \bibinfo{pages}{2564--2572}.
\newblock


\bibitem[Kearns et~al\mbox{.}(2019)]%
        {kearns2019empirical}
\bibfield{author}{\bibinfo{person}{Michael Kearns}, \bibinfo{person}{Seth Neel}, \bibinfo{person}{Aaron Roth}, {and} \bibinfo{person}{Zhiwei~Steven Wu}.} \bibinfo{year}{2019}\natexlab{}.
\newblock \showarticletitle{An empirical study of rich subgroup fairness for machine learning}. In \bibinfo{booktitle}{\emph{Proceedings of the conference on fairness, accountability, and transparency}}. \bibinfo{pages}{100--109}.
\newblock


\bibitem[Koren et~al\mbox{.}(2009)]%
        {koren2009matrix}
\bibfield{author}{\bibinfo{person}{Yehuda Koren}, \bibinfo{person}{Robert Bell}, {and} \bibinfo{person}{Chris Volinsky}.} \bibinfo{year}{2009}\natexlab{}.
\newblock \showarticletitle{Matrix factorization techniques for recommender systems}.
\newblock \bibinfo{journal}{\emph{Computer}} \bibinfo{volume}{42}, \bibinfo{number}{8} (\bibinfo{year}{2009}), \bibinfo{pages}{30--37}.
\newblock


\bibitem[Kroer and Peysakhovich(2019)]%
        {Kroer2019}
\bibfield{author}{\bibinfo{person}{Christian Kroer} {and} \bibinfo{person}{Alexander Peysakhovich}.} \bibinfo{year}{2019}\natexlab{}.
\newblock \bibinfo{title}{Scalable fair division for 'at most one' preferences}.
\newblock
\newblock
\showeprint[arxiv]{1909.10925}~[cs.GT]


\bibitem[Lee(2017)]%
        {Lee2017}
\bibfield{author}{\bibinfo{person}{Euiwoong Lee}.} \bibinfo{year}{2017}\natexlab{}.
\newblock \showarticletitle{{APX}-hardness of maximizing {Nash} social welfare with indivisible items}.
\newblock \bibinfo{journal}{\emph{Inform. Process. Lett.}}  \bibinfo{volume}{122} (\bibinfo{year}{2017}), \bibinfo{pages}{17--20}.
\newblock


\bibitem[Li et~al\mbox{.}(2022)]%
        {Li2022Fairness}
\bibfield{author}{\bibinfo{person}{Yunqi Li}, \bibinfo{person}{Hanxiong Chen}, \bibinfo{person}{Shuyuan Xu}, \bibinfo{person}{Yingqiang Ge}, \bibinfo{person}{Juntao Tan}, \bibinfo{person}{Shuchang Liu}, {and} \bibinfo{person}{Yongfeng Zhang}.} \bibinfo{year}{2022}\natexlab{}.
\newblock \showarticletitle{Fairness in recommendation: a Survey}.
\newblock \bibinfo{journal}{\emph{arXiv preprint arXiv:2205.13619}} (\bibinfo{year}{2022}).
\newblock


\bibitem[Lipton et~al\mbox{.}(2004)]%
        {Lipton2004}
\bibfield{author}{\bibinfo{person}{Richard~J Lipton}, \bibinfo{person}{Evangelos Markakis}, \bibinfo{person}{Elchanan Mossel}, {and} \bibinfo{person}{Amin Saberi}.} \bibinfo{year}{2004}\natexlab{}.
\newblock \showarticletitle{On approximately fair allocations of indivisible goods}. In \bibinfo{booktitle}{\emph{Proceedings of the 5th ACM Conference on Electronic Commerce}}. \bibinfo{pages}{125--131}.
\newblock


\bibitem[Mine et~al\mbox{.}(2013)]%
        {mine2013reciprocal}
\bibfield{author}{\bibinfo{person}{Tsunenori Mine}, \bibinfo{person}{Tomoyuki Kakuta}, {and} \bibinfo{person}{Akira Ono}.} \bibinfo{year}{2013}\natexlab{}.
\newblock \showarticletitle{Reciprocal recommendation for job matching with bidirectional feedback}. In \bibinfo{booktitle}{\emph{Proceedings of the 2nd IIAI International Conference on Advanced Applied Informatics}}. \bibinfo{pages}{39--44}.
\newblock


\bibitem[Neve and Palomares(2019a)]%
        {neve2019aggregation}
\bibfield{author}{\bibinfo{person}{James Neve} {and} \bibinfo{person}{Ivan Palomares}.} \bibinfo{year}{2019}\natexlab{a}.
\newblock \showarticletitle{Aggregation Strategies in User-to-User Reciprocal Recommender Systems}. In \bibinfo{booktitle}{\emph{2019 IEEE International Conference on Systems, Man and Cybernetics}}. \bibinfo{pages}{4031--4036}.
\newblock


\bibitem[Neve and Palomares(2019b)]%
        {neve2019latent}
\bibfield{author}{\bibinfo{person}{James Neve} {and} \bibinfo{person}{Ivan Palomares}.} \bibinfo{year}{2019}\natexlab{b}.
\newblock \showarticletitle{Latent factor models and aggregation operators for collaborative filtering in reciprocal recommender systems}. In \bibinfo{booktitle}{\emph{Proceedings of the 13th ACM Conference on Recommender Systems}}. \bibinfo{pages}{219--227}.
\newblock


\bibitem[Nguyen et~al\mbox{.}(2014)]%
        {Nguyen2014}
\bibfield{author}{\bibinfo{person}{Nhan-Tam Nguyen}, \bibinfo{person}{Trung~Thanh Nguyen}, \bibinfo{person}{Magnus Roos}, {and} \bibinfo{person}{J\"{o}rg Rothe}.} \bibinfo{year}{2014}\natexlab{}.
\newblock \showarticletitle{Computational complexity and approximability of social welfare optimization in multiagent resource allocation}.
\newblock \bibinfo{journal}{\emph{Autonomous Agents and Multi-Agent Systems}} \bibinfo{volume}{28}, \bibinfo{number}{2} (\bibinfo{year}{2014}), \bibinfo{pages}{256--289}.
\newblock


\bibitem[Palomares et~al\mbox{.}(2021)]%
        {palomares2021reciprocal}
\bibfield{author}{\bibinfo{person}{Iv{\'a}n Palomares}, \bibinfo{person}{Carlos Porcel}, \bibinfo{person}{Luiz Pizzato}, \bibinfo{person}{Ido Guy}, {and} \bibinfo{person}{Enrique Herrera-Viedma}.} \bibinfo{year}{2021}\natexlab{}.
\newblock \showarticletitle{Reciprocal recommender systems: analysis of state-of-art literature, challenges and opportunities towards social recommendation}.
\newblock \bibinfo{journal}{\emph{Information Fusion}}  \bibinfo{volume}{69} (\bibinfo{year}{2021}), \bibinfo{pages}{103--127}.
\newblock


\bibitem[Patro et~al\mbox{.}(2020)]%
        {patro2020fairrec}
\bibfield{author}{\bibinfo{person}{Gourab~K Patro}, \bibinfo{person}{Arpita Biswas}, \bibinfo{person}{Niloy Ganguly}, \bibinfo{person}{Krishna~P. Gummadi}, {and} \bibinfo{person}{Abhijnan Chakraborty}.} \bibinfo{year}{2020}\natexlab{}.
\newblock \showarticletitle{FairRec: two-sided fairness for personalized recommendations in two-sided platforms}. In \bibinfo{booktitle}{\emph{Proceedings of The Web Conference 2020}}. \bibinfo{pages}{1194--1204}.
\newblock


\bibitem[Pizzato et~al\mbox{.}(2010)]%
        {pizzato2010recon}
\bibfield{author}{\bibinfo{person}{Luiz Pizzato}, \bibinfo{person}{Tomek Rej}, \bibinfo{person}{Thomas Chung}, \bibinfo{person}{Irena Koprinska}, {and} \bibinfo{person}{Judy Kay}.} \bibinfo{year}{2010}\natexlab{}.
\newblock \showarticletitle{RECON: a reciprocal recommender for online dating}. In \bibinfo{booktitle}{\emph{Proceedings of the 5th ACM Conference on Recommender Systems}}. \bibinfo{pages}{207--214}.
\newblock


\bibitem[Saito and Joachims(2022)]%
        {SaitoJoachimsKDD2022}
\bibfield{author}{\bibinfo{person}{Yuta Saito} {and} \bibinfo{person}{Thorsten Joachims}.} \bibinfo{year}{2022}\natexlab{}.
\newblock \showarticletitle{Fair Ranking as Fair Division: Impact-Based Individual Fairness in Ranking}. In \bibinfo{booktitle}{\emph{Proceedings of the 28th ACM SIGKDD Conference on Knowledge Discovery and Data Mining}}. \bibinfo{pages}{1514--1524}.
\newblock


\bibitem[Singh and Joachims(2018)]%
        {singh2018fairness}
\bibfield{author}{\bibinfo{person}{Ashudeep Singh} {and} \bibinfo{person}{Thorsten Joachims}.} \bibinfo{year}{2018}\natexlab{}.
\newblock \showarticletitle{Fairness of exposure in rankings}. In \bibinfo{booktitle}{\emph{Proceedings of the 24th ACM SIGKDD International Conference on Knowledge Discovery and Data Mining}}. \bibinfo{pages}{2219--2228}.
\newblock


\bibitem[Steinhaus(1948)]%
        {steinhaus1948problem}
\bibfield{author}{\bibinfo{person}{Hugo Steinhaus}.} \bibinfo{year}{1948}\natexlab{}.
\newblock \showarticletitle{The problem of fair division}.
\newblock \bibinfo{journal}{\emph{Econometrica}} \bibinfo{volume}{16}, \bibinfo{number}{1} (\bibinfo{year}{1948}), \bibinfo{pages}{101--104}.
\newblock


\bibitem[Su et~al\mbox{.}(2022a)]%
        {su2022optimizing}
\bibfield{author}{\bibinfo{person}{Yi Su}, \bibinfo{person}{Magd Bayoumi}, {and} \bibinfo{person}{Thorsten Joachims}.} \bibinfo{year}{2022}\natexlab{a}.
\newblock \showarticletitle{Optimizing rankings for recommendation in matching markets}. In \bibinfo{booktitle}{\emph{Proceedings of the ACM Web Conference 2022}}. \bibinfo{pages}{328--338}.
\newblock


\bibitem[Su et~al\mbox{.}(2022b)]%
        {Su2022}
\bibfield{author}{\bibinfo{person}{Yi Su}, \bibinfo{person}{Magd Bayoumi}, {and} \bibinfo{person}{Thorsten Joachims}.} \bibinfo{year}{2022}\natexlab{b}.
\newblock \showarticletitle{Optimizing rankings for recommendation in matching markets}. In \bibinfo{booktitle}{\emph{Proceedings of the ACM Web Conference 2022}}. \bibinfo{pages}{328--338}.
\newblock


\bibitem[Tomita et~al\mbox{.}(2023)]%
        {Tomita2023}
\bibfield{author}{\bibinfo{person}{Yoji Tomita}, \bibinfo{person}{Riku Togashi}, \bibinfo{person}{Yuriko Hashizume}, {and} \bibinfo{person}{Naoto Ohsaka}.} \bibinfo{year}{2023}\natexlab{}.
\newblock \showarticletitle{Fast and examination-agnostic reciprocal recommendation in matching markets}. In \bibinfo{booktitle}{\emph{Proceedings of the 17th ACM Conference on Recommender Systems}}. \bibinfo{pages}{12--23}.
\newblock


\bibitem[Tomita et~al\mbox{.}(2022)]%
        {tomita2022matching}
\bibfield{author}{\bibinfo{person}{Yoji Tomita}, \bibinfo{person}{Riku Togashi}, {and} \bibinfo{person}{Daisuke Moriwaki}.} \bibinfo{year}{2022}\natexlab{}.
\newblock \showarticletitle{Matching theory-based recommender systems in online dating}. In \bibinfo{booktitle}{\emph{Proceedings of the 16th ACM Conference on Recommender Systems}}. \bibinfo{pages}{18--23}.
\newblock


\bibitem[Varian(1974)]%
        {Varian1974}
\bibfield{author}{\bibinfo{person}{Hal~R. Varian}.} \bibinfo{year}{1974}\natexlab{}.
\newblock \showarticletitle{Equity, envy, and efficiency}.
\newblock \bibinfo{journal}{\emph{Journal of Economic Theory}} \bibinfo{volume}{9}, \bibinfo{number}{1} (\bibinfo{year}{1974}), \bibinfo{pages}{63--91}.
\newblock


\bibitem[Wang and Chen(2021)]%
        {wang2021user}
\bibfield{author}{\bibinfo{person}{Ningxia Wang} {and} \bibinfo{person}{Li Chen}.} \bibinfo{year}{2021}\natexlab{}.
\newblock \showarticletitle{User bias in beyond-accuracy measurement of recommendation algorithms}. In \bibinfo{booktitle}{\emph{Proceedings of the 15th ACM Conference on Recommender Systems}}. \bibinfo{pages}{133--142}.
\newblock


\bibitem[Wang et~al\mbox{.}(2023)]%
        {Wang2023FairnessRecommender}
\bibfield{author}{\bibinfo{person}{Yifan Wang}, \bibinfo{person}{Weizhi Ma}, \bibinfo{person}{Min Zhang}, \bibinfo{person}{Yiqun Liu}, {and} \bibinfo{person}{Shaoping Ma}.} \bibinfo{year}{2023}\natexlab{}.
\newblock \showarticletitle{A survey on the fairness of recommender systems}.
\newblock \bibinfo{journal}{\emph{ACM Transactions on Information Systems}} \bibinfo{volume}{41}, \bibinfo{number}{3} (\bibinfo{year}{2023}), \bibinfo{pages}{1--43}.
\newblock


\bibitem[Wu et~al\mbox{.}(2021)]%
        {WuCaoXuTan2021}
\bibfield{author}{\bibinfo{person}{Yao Wu}, \bibinfo{person}{Jian Cao}, \bibinfo{person}{Guandong Xu}, {and} \bibinfo{person}{Yudong Tan}.} \bibinfo{year}{2021}\natexlab{}.
\newblock \showarticletitle{TFROM: a two-sided fairness-aware recommendation model for both customers and providers}. In \bibinfo{booktitle}{\emph{Proceedings of the 44th International ACM SIGIR Conference on Research and Development in Information Retrieval}}. \bibinfo{pages}{1013--1022}.
\newblock


\bibitem[Xia et~al\mbox{.}(2019)]%
        {xia2019we}
\bibfield{author}{\bibinfo{person}{Bin Xia}, \bibinfo{person}{Junjie Yin}, \bibinfo{person}{Jian Xu}, {and} \bibinfo{person}{Yun Li}.} \bibinfo{year}{2019}\natexlab{}.
\newblock \showarticletitle{WE-Rec: a fairness-aware reciprocal recommendation based on Walrasian equilibrium}.
\newblock \bibinfo{journal}{\emph{Knowledge-Based Systems}}  \bibinfo{volume}{182} (\bibinfo{year}{2019}), \bibinfo{pages}{104857}.
\newblock


\bibitem[Xia et~al\mbox{.}(2015)]%
        {xia2015reciprocal}
\bibfield{author}{\bibinfo{person}{Peng Xia}, \bibinfo{person}{Benyuan Liu}, \bibinfo{person}{Yizhou Sun}, {and} \bibinfo{person}{Cindy Chen}.} \bibinfo{year}{2015}\natexlab{}.
\newblock \showarticletitle{Reciprocal recommendation system for online dating}. In \bibinfo{booktitle}{\emph{Proceedings of the 2015 IEEE/ACM International Conference on Advances in Social Networks Analysis and Mining 2015}}. \bibinfo{pages}{234--241}.
\newblock


\bibitem[Zhu et~al\mbox{.}(2020)]%
        {zhu2020measuring}
\bibfield{author}{\bibinfo{person}{Ziwei Zhu}, \bibinfo{person}{Jianling Wang}, {and} \bibinfo{person}{James Caverlee}.} \bibinfo{year}{2020}\natexlab{}.
\newblock \showarticletitle{Measuring and mitigating item under-recommendation bias in personalized ranking systems}. In \bibinfo{booktitle}{\emph{Proceedings of the 43rd international ACM SIGIR conference on research and development in information retrieval}}. \bibinfo{pages}{449--458}.
\newblock


\end{thebibliography}

\end{document}